\documentstyle[preprint,prd,aps]{revtex}

\tighten

\begin{document}

\preprint{TRI-PP-93-88 ---
SFU HEP-112-93 --- hep-lat/9312008 --- Revised August 1994}

\title{Gauge fixing and extended Abelian monopoles \\
in SU(2) gauge theory in $2+1$ dimensions}

\author{G. I. Poulis}
\address{TRIUMF, Theory Group,
4004 Wesbrook Mall, Vancouver, B.C. Canada V6T 2A3}

\author{Howard D. Trottier}
\address{Department of Physics, Simon Fraser University,
Burnaby, B.C. Canada V5A 1S6}

\author{R. M. Woloshyn}
\address{TRIUMF, Theory Group,
4004 Wesbrook Mall, Vancouver, B.C. Canada V6T 2A3}


\maketitle

\begin{abstract}
Extended Abelian monopoles are investigated in SU(2) lattice gauge 
theory in three dimensions. Monopoles are computed by Abelian 
projection in several gauges, including the maximal 
Abelian gauge. The number $N_m$ of extended monopoles in a cube 
of size $m^3$ (in lattice units)
is defined as the number of elementary ($1^3$) monopoles
minus antimonopoles in the cube ($m=1,2,\ldots$).
The distribution of $1^3$ monopoles in the nonlocal maximal Abelian 
gauge is shown to be essentially random, while
nonscaling of the density of $1^3$ monopoles
in some local gauges, which has been previously observed, 
is shown to be mainly due to strong short-distance correlations.
The density of extended monopoles in local gauges
is studied as a function of $\beta$ for monopoles of
fixed physical ``size'' ($m / \beta = {\rm fixed}$);
the degree of scale violation is found to decrease 
substantially as the monopole size is increased.
The possibility therefore remains that long distance 
properties of monopoles in local gauges may be relevant 
to continuum physics, such as confinement.
\end{abstract}
   

\section{Introduction} 
The dual Meissner effect was suggested as possible mechanism for 
confinement in nonAbelian gauge theories more than a decade ago by 
Mandelstam \cite{Mandel} and t' Hooft \cite{HooftFirst,Hooft}, and 
has recently been the subject of intensive lattice investigations
\cite{KronNP,KronPL,Green,IvanExtend,Suzuki,Born4D,DebbioF12,
Hioki,Born3D,IvanTc,DualAbrikosov,Stack,DiGiacomoTc,SuzukiTc}. 
In this picture of confinement, (color) magnetic monopoles are 
presumed to condense in the vacuum, forcing (color) electric fields 
between two sources to be squeezed into a flux tube. Magnetic 
monopoles are well understood to result in confinement in compact 
Quantum Electrodynamics, as has been demonstrated both 
analytically \cite{Poly,Banks} and numerically \cite{DeGrand,QED3}. 
The conjecture is that such ``Abelian'' degrees of freedom
also drive confinement in nonAbelian theories. 't Hooft proposed 
that this mechanism can be realized in nonAbelian theories
by imposing a gauge fixing condition that is invariant under a 
Cartan subgroup U(1)${}^{N-1}$ of the original SU($N$) theory,
referred to as an Abelian projection \cite{Hooft}.
Singularities in the gauge-fixing condition are identified with 
monopole world lines in four dimensions ($d=4$) and monopole points 
(instantons) in three dimensions ($d=3$).

A lattice implementation of Abelian projection was
formulated in Ref. \cite{KronNP,KronPL}, in which
several gauge-fixing conditions were also developed 
(following `t~Hooft \cite{Hooft}) that have been
widely studied. The so-called maximal Abelian gauge 
is defined by gauge transformations $G$ which
maximize the following quantity \cite{KronPL}
(we henceforth restrict our attention to SU(2)):
\begin{equation}
   R \equiv \sum_{x,\mu} {\rm Tr} \Bigl( 
   \sigma_3 G(x) U(x,\hat\mu) G^\dagger(x+\hat\mu) 
   \sigma_3 G(x) U^\dagger(x,\hat\mu) G^\dagger(x+\hat\mu) \Bigr) ,
\label{MAgauge}
\end{equation}
where $U(x,\hat\mu)$ are link variables.
In the continuum limit, ${\rm max}(R)$ reduces
to the renormalizable differential gauge
$(\partial_{\mu}\pm igA_{\mu}^{3})A^{\pm,\mu}=0$,
where $A^\pm_\mu \equiv (A^1_\mu \pm i A^2_\mu) / \sqrt2$.

Local (generally nonrenormalizable) gauges can be defined by the 
diagonalization of an adjoint operator $\Phi$ \cite{KronNP}:
\begin{equation}
   \Phi(x) \rightarrow G(x) \Phi(x) G^{\dagger}(x)
   = \left( \begin{array}{cc}
   e^{i\alpha(x)} & 0 \\     
   0 & e^{-i\alpha(x)}
   \end{array} \right) .
\label{local}
\end{equation}
Examples of Eq. (\ref{local}) are diagonalization of
a plaquette or a Polyakov line \cite{KronNP}. 
We will refer to a gauge as ``local'' when the gauge 
condition can be imposed on a site-by-site basis.

Notice that Eqs. (\ref{MAgauge}) and (\ref{local}) are invariant
under a local U(1) transformation 
$G\rightarrow d G$, where $d$ is a diagonal matrix 
$d=\cos\theta+i{\sigma}_3\sin\theta$ \cite{Residual}.
The Abelian projection of a link $U$ is defined, after
gauge-fixing, as its component $u$ in the corresponding
subspace \cite{KronNP,KronPL}
\begin{equation}
   G(x) U(x,\hat\mu) G^\dagger(x+\hat\mu)
   \equiv w(x,\hat\mu) \, u(x,\hat\mu) ,
   \quad {\rm Tr}(\sigma_3 w) \equiv 0
\label{Project}
\end{equation} 
($w^\dagger w \equiv u^\dagger u \equiv 1$). 
The Abelian projection $u(x,\hat\mu)$ can be written as 
$u(x,\hat\mu) = \exp[i \sigma_3 \phi(x,\hat\mu)]$.
Monopoles are defined from the phases $\phi$
(as described in Sec.~II) and are associated with 
singularities in the gauge fixing which, in the context of 
Eq. (\ref{local}), occur at points where
$\alpha(x) = 0$ \cite{Hooft}. 

One of the main issues that has been addressed
in lattice investigations is whether the Abelian monopole
density in a particular gauge exhibits scaling
(thus having a well defined continuum limit).
SU(2) lattice simulations in maximal Abelian gauge 
show scaling behavior for monopoles defined on elementary
cubes of the lattice in $d=3$ \cite{Born3D},
and there is some evidence for scaling in 
$d=4$ \cite{Born4D,DebbioF12}.
On the other hand, it has been well established that the
density of elementary monopoles in a variety of local gauges
does {\it not\/} scale (this includes diagonalization of a 
plaquette and the Polyakov loop) \cite{DebbioF12}. 

As a result, the maximal Abelian gauge has become widely 
regarded as the only known gauge which may yet establish
the role (if any) of Abelian monopoles in confinement.
However, there is as yet no compelling reason why the 
degrees of freedom responsible for confinement in a 
gauge-invariant theory should only be manifest in 
this particular gauge. In fact, some monopole operators
have recently been constructed which show evidence
in local gauges for a spontaneous symmetry breaking 
that is correlated with the deconfinement phase transition
\cite{DiGiacomoTc,SuzukiTc}.

We reconsider the lattice scale dependence of the monopole
density in local gauges by analyzing the properties of 
{\it extended\/} monopoles.
We work in three-dimensional SU(2) lattice gauge theory.
The number $N_m$ of extended monopoles in a cube of size $m^3$
(in lattice units) is defined as the number of elementary 
($1^3$) monopoles minus antimonopoles in the cube ($m=1,2,\ldots$). 
A physical reason for studying extended monopoles 
is that confinement occurs on 
some finite physical length scale. The nonscaling behavior 
that has been observed in some gauges for elementary monopoles
might therefore not be a good criterion for ruling out
contact with the physics relevant to confinement. 

Indeed, we find that nonscaling of the monopole density 
in some local gauges is due
mainly to short-distance monopole-antimonopole fluctuations, 
which diverge in number as the coupling $\beta \to \infty$.
On the other hand, such fluctuations are essentially absent 
in the maximal Abelian gauge. This is reflected in our calculation 
of the average minimum separation between a monopole and 
the nearest neighboring antimonopole, $\langle r_{\rm min} \rangle$.
In maximal Abelian gauge $\langle r_{\rm min} \rangle$ 
scales with $\beta$, while it vanishes (in physical units)
with increasing $\beta$ in the local gauges that we considered.
[The dimensionless quantity $\beta = 4/(g^2a)$, 
where the coupling constant $g$ has units of 
$({\rm mass})^{1/2}$ in three dimensions.
We work throughout in lattice units, where
the spacing $a\equiv1$.] 

Correlations between elementary monopoles and 
antimonopoles as functions of their separation are also calculated.
Strong short-distance correlations are found in local gauges.
On the other hand, in maximal Abelian gauge the monopole distribution
is essentially random (the monopoles form a ``plasma'').
We speculate that large short-distance fluctuations in
the gauge fields lead to the divergence in the number of 
monopoles in local gauges, while in the maximal Abelian gauge,
which is imposed in nonlocal way, these fluctuations are 
effectively smoothed out.

We then consider the density of extended monopoles.
In local gauges the density is found to decrease 
rapidly as the monopole ``size'' $m$ is increased, the rate 
of change being maximal for $m$ near unity. This can
again be understood as due to an averaging over
short distance fluctuations (i.e., as an averaging of 
the charges of elementary monopole-antimonopole pairs over
larger volumes). By contrast, the density varies 
slowly with $m$ in maximal Abelian gauge. 
We also consider the density of extended monopoles
of fixed physical ``size'' ($m / \beta = {\rm fixed}$),
as a function of $\beta$. For a finite value of
$m / \beta$ the density of monopoles in a typical local
gauge does not scale; however, we find that the degree of 
scaling violation decreases substantially as 
the physical monopole ``size'' increases.

These results support the argument that strong short-distance 
fluctuations in the gauge fields can be effectively
averaged out by a nonlocal gauge condition, as in the
maximal Abelian gauge, or (to a large extent) 
by considering extended monopoles on sufficiently 
long length scales in local gauges. 

The organization of the rest of this paper is as follows. 
In Sec.~II the various gauge fixing schemes and observables 
considered in this work are described.
Our results are presented in Sec.~III,
and in Sec.~IV we give our conclusions.

\section{Method}

Abelian projection was carried out in the maximal Abelian
gauge and in a number of local gauges.
The maximization of $R$, Eq. (\ref{MAgauge}), 
which defines the maximal Abelian  gauge must be implemented
iteratively. Following Ref. \cite{Born4D} we repeatedly
sweep through the lattice, maximizing $R$ locally by 
solving for $G(x)$ analytically at each site, 
keeping $G(x+\hat\mu)$ at neighboring sites fixed. 
These iterations are repeated until $G$ at all sites becomes
sufficiently close to the identity:
\begin{equation}
   {\rm max} \{ 1- {\case12} {\rm Tr}\,G(x) \} \le \delta \ll 1 ,
\label{crit}   
\end{equation}
with $\delta = O(10^{-7})$ used as a stopping criterion. 
Note that this iteration 
procedure is not guaranteed to reach the the global maximum of 
$R$ \cite{Born4D}. Moreover, the iteration procedure 
is not guaranteed to increase ${\rm Tr}\,G(x)$ monotonically over 
successive iterations; hence the stopping criterion Eq. (\ref{crit}) 
is ambiguous. As a result, in practice the monopole number 
is not uniquely defined in the maximal Abelian gauge.
For example, two field configurations differing only by an SU(2) 
gauge transformation will not, in practice, necessarily lead to the 
same number of Abelian monopoles in an actual calculation 
in maximal Abelian gauge. The implementation of the maximal 
Abelian gauge fixing also dominates
the computational cost of the simulation, due to the large
number of iterations (typically 1000) required to satisfy
Eq. (\ref{crit}).
             
A local gauge that we consider is a simple implementation
of Eq. (\ref{local}): diagonalization of the field strength
tensor, $\Phi = F_{12}$ (cf. Ref. \cite{KronNP}).
Note that Abelian projection in this
gauge has the property that a pure-gauge configuration (a gauge
transformation of links equal to the identity) will in general
lead to a nonzero monopole density. Furthermore, the monopole 
numbers extracted from gauge-equivalent configurations in 
this gauge will not be equal in general. 
Calculations were also carried out in
gauges defined by the diagonalization of rectangular
Wilson loops of various sizes, as well as in an
``unfixed'' gauge (defined by $\Phi=1$ in Eq. (\ref{local})).
These calculations lead to the same conclusions
as drawn from the results presented here for 
$\Phi=F_{12}$.

Following Ref. \cite{KronNP} we compute phases of the
Abelian projections of the gauge-fixed links 
$u(x,\mu) \equiv u_0(x,\mu) + i \vec\sigma \cdot \vec u(x,\mu)$
(cf. Eq. (\ref{Project})):
\begin{equation}\label{abel_pr}
   \phi(x,\hat\mu) = {\tan}^{-1}
   \left[u_3(x,\hat\mu) \over u_0(x,\hat\mu) \right],
   \quad \phi(x,\hat\mu) \in (-\pi,+\pi] .
\end{equation}
Reduced plaquette angles $\widetilde\phi$ are
then defined according to \cite{DeGrand}: 
\begin{equation}
   \widetilde\phi \equiv
   \phi - 2 \pi N_s , 
   \quad \widetilde\phi \in (-\pi, +\pi] ,
\label{plaqangle}
\end{equation}
where $N_s$ is identified with the number of Dirac strings 
passing through the plaquette 
($N_s \in [-2,2]$). The number of ``elementary'' 
monopoles $N_{m=1}$ contained in a cube of size $1^3$ is equal to
the sum of Dirac strings $N_s$ passing through the oriented 
$1\times 1$ plaquettes on the surfaces of the cube.

We define the number of extended monopoles
$N_m$ in a cube of size $m^3$ as the number of
elementary monopoles minus antimonopoles in the cube.
The density $\rho_m$ of extended monopoles of
``size'' $m$ is given by
\begin{equation}
   \rho_m = {1\over 2 L^3} \sum_{i=1}^{(L/m)^3} \vert N_m(x_i) \vert ,
\end{equation}
where $x_i$ labels the coordinates of the cubes of size $m^3$.
We take $m$ to divide the lattice length $L$ (although
this requirement can be relaxed); hence the total number of extended 
monopoles on the lattice vanishes.

The definition of $N_m$ above corresponds to ``type-II'' extended 
monopoles defined in Ref. \cite{IvanExtend};
a ``type-I'' extended monopole number was also defined, 
which is computed from the phases of $m\times m$ plaquettes on
the surface of a cube of size $m^3$ 
(using a suitable generalization of Eq. (\ref{plaqangle})).
In an earlier version of this work we considered type-I extended 
monopoles. However, for large $m$ the number of type-I monopoles 
approaches the strong coupling limit, since the phases of the 
$m\times m$ plaquettes become essentially random \cite{Cappelli,KYee}.
Since we are interested in weak coupling continuum physics
we consider only type-II monopoles in the rest of this paper. 

The simulations were done mainly on a $24^3$ lattice, using a
bath algorithm at several values of $\beta$ in the scaling region.
The string tension on lattices of this size is found
to scale for $\beta \gtrsim 6$ (see, e.g., 
Ref. \cite{MawhScaling}). Some data was also taken at 
smaller values of $\beta$ for comparison.
We find that finite volume effects on the $24^3$ lattice, 
as measured by the expectation value of the Polyakov line,
become noticeable for $\beta \gtrsim 12$ (which is the 
largest value in our data). 
Measurements were made on an ensemble
of 500 configurations each separated by 100 updates,
which is significantly longer than the autocorrelation time 
for any observable that we considered (the maximal Abelian gauge
exhibits the longest autocorrelation time of the various
gauges). We also used 200 configurations on a $36^3$ lattice
to obtain some results at large $m$
(these data are identified explicitly in the following).

\section{Results}

The density $\rho_1$ of elementary ($m=1$) monopoles is
shown in Fig.~1 in maximal Abelian  gauge 
(cf. Eq. (\ref{MAgauge})) and $F_{12}$-gauge 
($\Phi=F_{12}$ in Eq. (\ref{local})).
It is clear that only 
maximal Abelian gauge exhibits scaling in $\rho_1$, and that 
the difference between the two gauges becomes more 
pronounced as $\beta$ is increased. 

Notice that maximal Abelian
gauge has far fewer monopoles than the local gauges.
To help understand this difference it is instructive to look
at the spatial distribution of monopoles and antimonopoles. 
This is done by counting the number of monopoles $N(r_{\rm min})$
for which the nearest antimonopole is a distance $r_{\rm min}$
away. The results are shown in Fig.~2 for the $F_{12}$-gauge.
As expected with a relatively
large number of monopoles and antimonopoles, there are many short
distance pairs; in fact, most monopoles have an antimonopole
in a neighboring lattice cell.
A natural question is whether the monopoles and
antimonopoles are correlated. Some insight into this question is
obtained by comparing the result of the lattice simulation with 
$N(r_{\rm min})$ for a completely random distribution of monopoles
and antimonopoles. This is also shown in Fig.~2. A random distribution
leads to an $N(r_{\rm min})$ which is much broader than that observed
in $F_{12}$-gauge. This is indicative of a
strong short-range correlation between the monopoles and antimonopoles
in this gauge, with significant enhancement of nearest-neighbor pairs.

Figure 3 shows $N(r_{\rm min})$ for the maximal Abelian gauge 
at $\beta = 8$. Here the number of monopoles 
with nearby antimonopoles is very small
so statistical errors are large, but $N(r_{\rm min})$ is roughly
consistent with a random distribution. There certainly is no 
enhancement of pairs with small separations as in $F_{12}$-gauge.

Figure 4 shows the average value of the minimum separation of a
monopole and antimonopole $\langle r_{\rm min} \rangle$ for the
two gauges as function of the number of monopoles (corresponding
to $\beta$ in the range 5 to 12). 
Again, for comparison, the results expected
for a completely random distribution are plotted. For monopoles
in maximal Abelian gauge we find reasonable consistency with a
random distribution (a monopole ``plasma'') but in $F_{12}$-gauge
significantly smaller values of $\langle r_{\rm min} \rangle$
are observed.

If the monopole density scales and the distribution is random we 
certainly expect $\langle r_{\rm min} \rangle$ 
to scale with $\beta$. This is shown in Fig.~5. For maximal 
Abelian gauge, $\langle r_{\rm min} \rangle$ scales very well.
In $F_{12}$-gauge $\langle r_{\rm min} \rangle$ vanishes 
in physical units as $\beta$ increases.

These results support the idea that the excess of
monopoles found in local gauges relative to maximal Abelian gauge
is due to highly-correlated short distance fluctuations. 
This is further illustrated by considering the density $\rho_m$ 
of extended monopoles.

In Fig.~6 we compare the density $\rho_m$ in the two gauges
at $\beta=6$. The density in maximal Abelian gauge decreases
slowly with $m$; the decrease becomes more pronounced
for $m$ around the value of $\langle r_{\rm min} \rangle$ 
for $1^3$ monopoles in that gauge. This is consistent with 
the fact that strong short-distance correlations are absent in 
maximal Abelian gauge, with $1^3$ monopoles forming a plasma, 
distributed on a length scale of $O(\langle r_{\rm min} \rangle$).
By contrast, the density in $F_{12}$-gauge decreases
rapidly with $m$, with the rate of change being maximal near $m=1$.

Figure~7(a) shows that the density of extended monopoles 
of fixed physical ``size'' ($m / \beta = {\rm fixed}$) 
scales well in maximal Abelian gauge. 
In $F_{12}$-gauge the density does not scale for 
any finite value of $m/\beta$, as illustrated in Fig.~7(b). 
The density $\rho \beta^3$ in physical units increases roughly
linearly with $\beta$ for fixed $ m/\beta$.
However, the degree of scale violation decreases 
substantially as the physical monopole ``size'' increases.
This is made evident by a plot of the
slope $\Delta(\rho\beta^3) / \Delta\beta$
as a function of $m/\beta$, given in Fig.~8.
This is again consistent with a cancellation of
short-distance monopole-antimonopole fluctuations.

\section{Conclusions}
The results presented here demonstrate that the 
nonscaling of the density of elementary ($1^3$) monopoles in 
a variety of local gauges, which has been previously observed, is 
due in large part to strong short distance correlations in 
those gauges. The distribution of $1^3$ monopoles in maximal 
Abelian gauge was shown to be essentially random.
The degree of scale violation in the density of 
extended monopoles in local gauges was found to decrease 
substantially as the monopole ``size'' is increased.
Our results support the argument that strong short-distance 
fluctuations in the gauge fields can be effectively
averaged out by a nonlocal gauge condition, as in the
maximal Abelian gauge, or (to a large extent) 
by considering extended monopoles on sufficiently 
long physical length scales in local gauges. 
The possibility therefore
remains that long distance properties of monopoles
in local gauges may be relevant to continuum physics,
such as confinement.

\acknowledgments
We thank K. Yee for fruitful conversations. This work
was supported in part by the Natural Sciences and 
Engineering Research Council of Canada.


\begin{figure}
\caption{Density of elementary ($m=1$) monopoles $\rho_1 \beta^3$
in physical units as a function of $\beta$ in the maximal 
Abelian gauge (full circles) and $F_{12}$-gauge (full triangles).}
\end{figure}

\begin{figure}
\caption{The number of elementary monopoles $N(r_{\rm min})$ 
versus $r_{\rm min}$ at $\beta = 8$ in $F_{12}$-gauge 
(full triangles). Also shown is $N(r_{\rm min})$ for a 
completely random distribution (open circles).}
\end{figure}

\begin{figure}
\caption{The number of elementary monopoles $N(r_{\rm min})$ 
versus $r_{\rm min}$ at $\beta = 8$ in maximal Abelian gauge
(full circles). Also shown is $N(r_{\rm min})$ for a 
completely random distribution (open circles).}
\end{figure}

\begin{figure}
\caption{The average minimum monopole--antimonopole separation 
$\langle r_{\rm min} \rangle$ versus the number of $m=1$ monopoles 
$N_1$ in maximal Abelian gauge (full circles) and 
$F_{12}$-gauge (full triangles).
Also shown is $\langle r_{\rm min} \rangle$ for a completely 
random distribution (solid line).}
\end{figure}

\begin{figure}
\caption{The average minimum monopole--antimonopole separation
$\langle r_{\rm min} \rangle$ as a function of $\beta$, for 
$m=1$ monopoles in the maximal Abelian gauge (main figure)
and in the $F_{12}$-gauge (inset). The solid line in the 
inset shows a fit 
$\langle r_{\rm min} \rangle / \beta = c / \beta$
to the data in $F_{12}$-gauge.}
\end{figure}

\begin{figure}
\caption{Extended monopole density $\rho_m$ at $\beta=6$ 
as a function of the ``size'' $m$ in the maximal Abelian gauge 
(full circles) and $F_{12}$-gauge (full triangles).}
\end{figure}

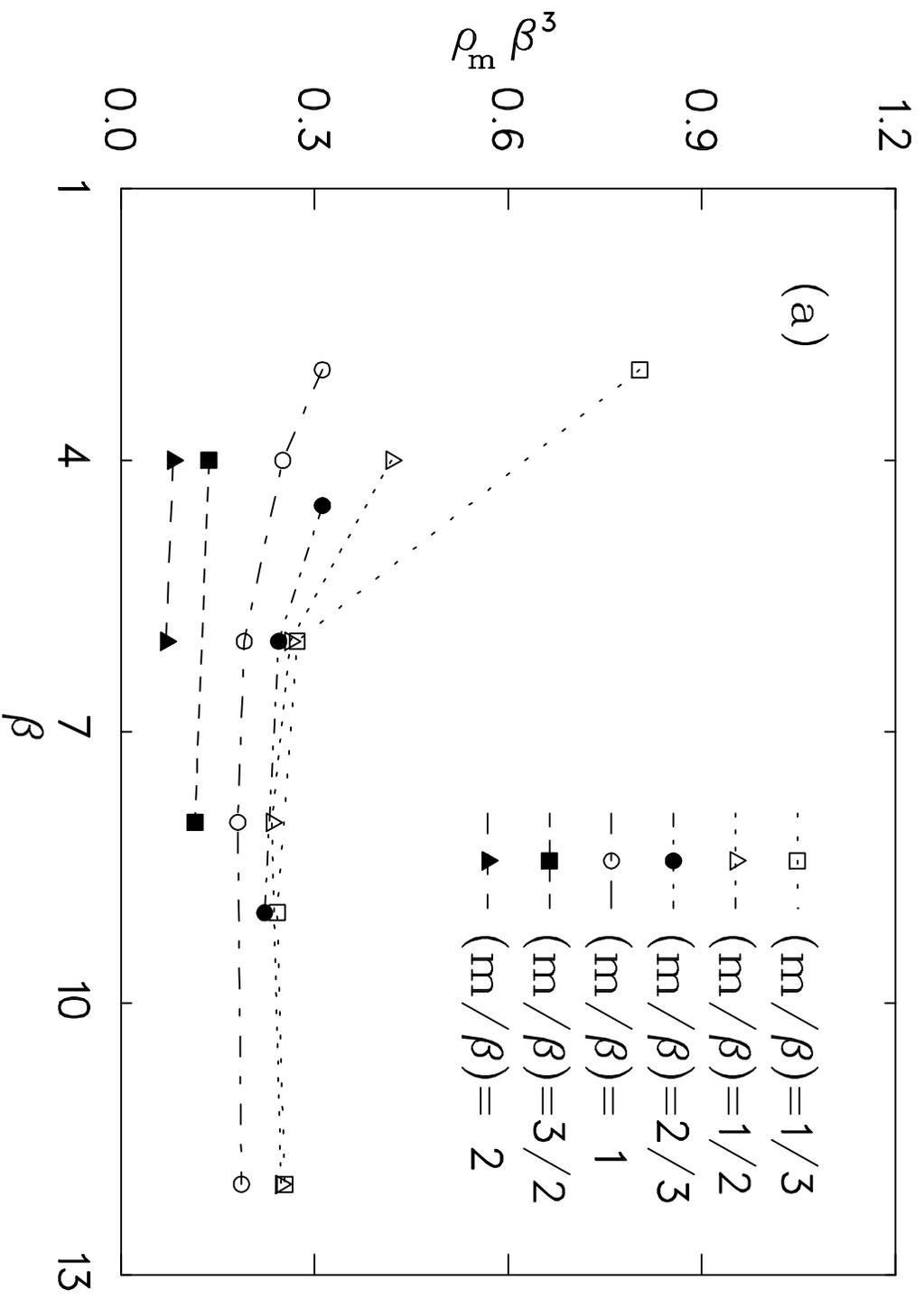
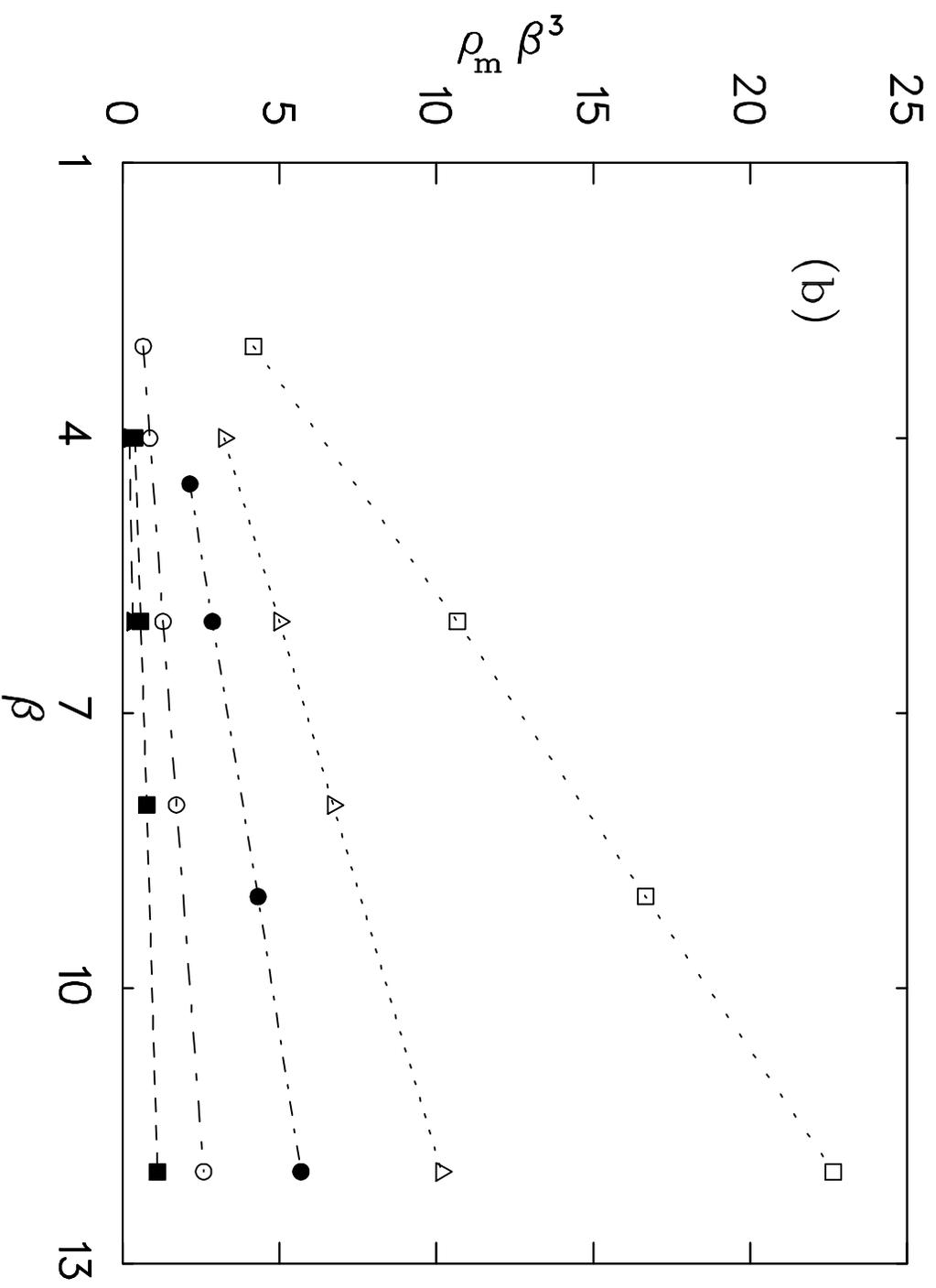
\begin{figure}
\caption{Extended monopole density $\rho$
as a function of $\beta$, for several fixed monopole 
``sizes'' in physical units ($m/\beta = {\rm fixed}$),
in (a) maximal Abelian gauge and (b) $F_{12}$-gauge.
The results for $F_{12}$ gauge are taken from
a $36^3$ lattice. Straight lines are shown in
Fig. 7(b) to guide the eye.}

\end{figure}

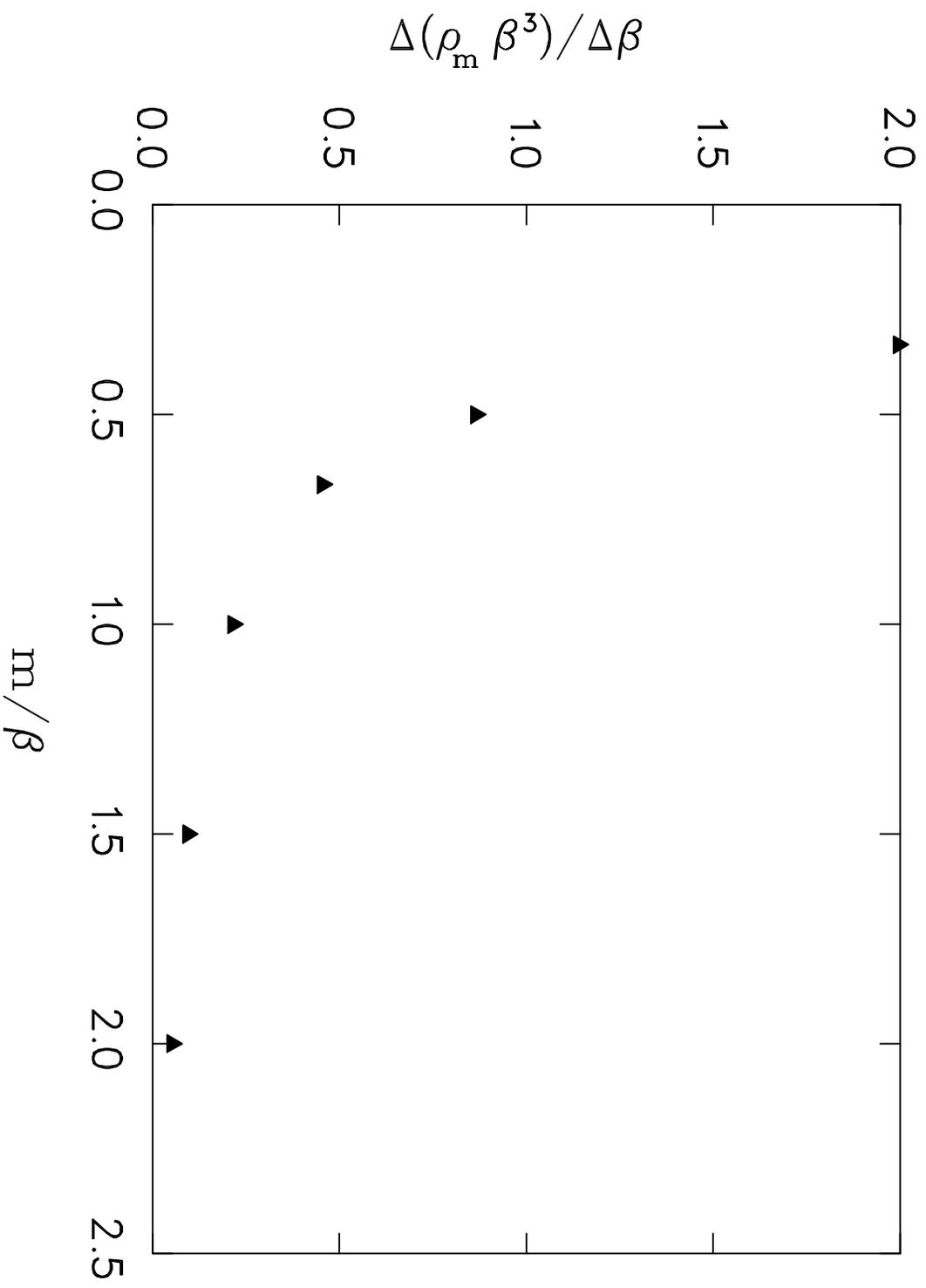
\begin{figure}
\caption{Slope $\Delta(\rho\beta^3) / \Delta\beta$
of the extended monopole density in $F_{12}$-gauge,
as a function of $m / \beta$. The slope is
estimated from the data in Fig.~7(b) using the 
two highest available $\beta$ values.}
\end{figure}


\begin{references}
\bibitem{Mandel}S. Mandelstam, Phys. Rep. {\bf 23C},
245 (1976); {\it Monopoles in Quantum Field Theory},
Proceedings of the Monopole Meeting, Trieste 1981, 
Eds. N. S. Craigie, P. Goddard and W. Nahm 
(World Scientific, Singapore, 1982).

\bibitem{HooftFirst}G. t'Hooft, Proceedings of the 
EPS International Conference on High Energy Physics, 
Palermo 1975, Ed. A. Zichichi 
(Editrice Compositori, Bologna, 1976).

\bibitem{Hooft}G. t'Hooft, Nucl. Phys. {\bf B190}, 455 (1981).  

\bibitem{KronNP}A. S. Kronfeld, G. Schierholz and U.-J. Wiese, 
Nucl. Phys. {\bf B293}, 461 (1987).

\bibitem{KronPL}A. S. Kronfeld, M. L. Laursen, G. Schierholz,
U.J. Wiese, Phys. Lett. {\bf B198}, 516 (1987).

\bibitem{Green}J. Greensite and J. Winchester,
Phys. Rev. D {\bf 40}, 4167 (1989).

\bibitem{IvanExtend}T.L. Ivanenko, A.V. Pochinsky and 
M.I. Polykarpov, Phys. Lett. {\bf B252}, 631 (1990).

\bibitem{Suzuki} T. Suzuki and I. Yotsuyanagi, 
Phys. Rev. D {\bf 42}, 4257 (1990).   

\bibitem{Born4D}V. G. Bornyakov, E. M. Ilgenfritz, 
M. L. Laursen, V. K. Mitryushkin, M. Muller-Preussker,
A.J. van der Sijs, A.M. Zadorozhnyi,
Phys. Lett. {\bf B261}, 116 (1991).

\bibitem{DebbioF12}L. Del Debbio, A. Di Giacomo
M. Maggiore, S. Olejnik, Phys. Lett. {\bf B267}, 254 (1991).  

\bibitem{Hioki}S. Hioki, S. Kitahara, S. Kiura, Y. Matsubara,
O. Miyamura, S. Ohno, T. Suzuki, 
Phys. Lett. {\bf B272}, 326 (1991). 

\bibitem{Born3D}V.G. Bornyakov and R. Grygoryev, 
Nucl. Phys. B (Proc. Suppl.) {\bf 30}, 576 (1993).    

\bibitem{IvanTc}T.L. Ivanenko, A.V. Pochinsky and 
M.I. Polykarpov, Phys. Lett. {\bf B302}, 458 (1993).

\bibitem{DualAbrikosov}V. Singh, D. Browne,
and R. W. Haymaker, Phys. Lett. {\bf B306}, 115 (1993);
P. Cea and L. Cosmai, Nuovo Cim. {\bf 107A}, 541 (1994).

\bibitem{Stack}J. D. Stack, S. D. Nieman, and R. J. Wensley,
preprint ILL-TH-94-14, 1994.

\bibitem{DiGiacomoTc}L. Del Debbio, A. Di Giacomo, G. Paffuti,
and P. Pieri, preprint IFUP-TH-30-94, 1994.

\bibitem{SuzukiTc}See also T. Suzuki, S. Ilyar, 
Y. Matsubara, T. Okude, K. Yotsuji,
preprint KANAZAWA-94-15, 1994.

\bibitem{Poly}A. M. Polyakov, 
Nucl. Phys. {\bf B120}, 429 (1977).

\bibitem{Banks}T. Banks, R. Myerson and J. Kogut, 
Nucl. Phys. {\bf B129}, 493 (1977).   

\bibitem{DeGrand}T. A. DeGrand and D. Toussaint, 
Phys. Rev. D {\bf 22}, 2478 (1980).

\bibitem{QED3}A. Irb\"ack and C. Peterson, 
Phys. Rev. D {\bf 36}, 3804 (1987);
R. J. Wensley and J. D. Stack, 
Phys. Rev. Lett. {\bf 63}, 1764 (1989).
See also H. D. Trottier and R. M. Woloshyn,
Phys. Rev. D {\bf 48}, 4450 (1993).

\bibitem{Residual}We observe that Eq. (\ref{MAgauge}) is also
invariant under a purely ``off-diagonal'' transformation
$G(x) \to N(x)G(x)$, where
$N(x) = i[\cos\theta(x) \sigma_1 + \sin\theta(x)\sigma_2]$
and $\theta$ is arbitrary.
The local gauge condition Eq. (\ref{local}) is invariant
under $G(x) \to \varphi(x) G(x)$, 
where $\varphi$ satisfies $[\varphi(x), \Phi(x)] = 0$.

\bibitem{Cappelli}A. Cappelli, Nucl. Phys. {\bf B275}, 488 (1986).

\bibitem{KYee}K. Yee, preprint LSUHEP-010194, 1994.
We thank K. Yee for bringing this argument to our attention. 

\bibitem{MawhScaling}R. D. Mawhinney, 
Phys. Rev. D {\bf 41}, 3209 (1990).

\end{references}
\end{document}